\documentclass[pre,twocolumn]{revtex4}

\usepackage[dvips]{graphicx}
\usepackage{amsmath}

\DeclareMathOperator{\ath}{tanh^{-1}}

\begin{document}

\title{The Complexity of Ising Spin Glasses}
\date{\today}
\author{T.\ Aspelmeier}
\author{A.\ J.\ Bray}
\author{M.\ A.\ Moore}
\affiliation{Department of Physics and Astronomy, University of Manchester,
Manchester M13 9PL, U.K.}

\begin{abstract}
We compute the complexity (logarithm of the number of TAP states) 
associated with minima and index-one saddle points of the TAP free 
energy. Higher-index saddles have smaller complexities. 
The two leading complexities are equal, consistent with the Morse 
theorem on the total number of turning points, and have the value given 
in [A. J. Bray and M. A. Moore, J. Phys. C {\bf 13}, L469 (1980)].  
In the thermodynamic limit, TAP states of all free energies become 
marginally stable.  

\end{abstract} 

\maketitle

The  computation  of  the  complexity  of spin  glasses  has  recently
attracted an avalanche of renewed interest. More than two decades ago,
two of  us \cite{BM80} (BM) computed  the average  number of
solutions, $\langle N_s \rangle_J$, of the TAP equations \cite{TAP} for
the SK Ising spin-glass model.  We found  $\langle N_s  \rangle_J \sim
\exp[N\Sigma(T)]$, where  $\Sigma(T)$ is the complexity  (per spin) at
temperature  $T$  and  $N$  is  the number  of  spins.   The  function
$\Sigma(T)$ vanishes at the spin-glass critical point $T_c$, while for
$T  \to   0$  it  approaches  the  value   $0.1992...$  obtained  from
independent  (and   much  simpler)  calculations  of   the  number  of
1-spin-flip-stable states at $T=0$ \cite{BM80,TE}.

Various aspects of this  calculation have recently been (implicitly or
explicitly)   criticized  \cite{CGPM,ACGP,CLPR1,ACGPT,CLPR2,CLR}.  The
essence of  this criticism is as  follows. The TAP  equations take the
form $G_i=0$,  $i=1,\ldots,N$, where each  $G_i$ is a function  of the
$N$   variables   $m_i$  which   denote   the  local   magnetizations,
$m_i=\langle  S_i  \rangle$, at  each  site,  where  the brackets  are
thermal averages  and $S_i =  \pm 1$ is  an Ising spin  variable.  The
number of  solutions of the $N$  equations $G_i=0$ is given  by $N_s =
\int_{-1}^1     \prod_{i}dm_i     \prod_i\delta(G_i)    |\det{\partial
G_i}/\partial m_j|$, where the modulus sign on the determinant ensures
that  each solution is  counted with  weight unity.   In BM  (and many
subsequent calculations) the modulus  sign was dropped.  This is valid
if the matrix $\partial  G_i/\partial m_j$ is positive definite, which
requires all TAP solutions to be  local minima of the TAP free energy.
We will  argue that, with  one modification, this idea  is essentially
correct.   This seems,  at first  sight, paradoxical  because,  if the
modulus sign is dropped, TAP solution $s$ is weighted by $(-1)^{n_s}$,
where  $n_s$ is  the index  (number  of negative  eigenvalues) of  the
saddle  point, giving $N_s  = \sum_s  (-1)^{n_s}$.  The  Morse theorem
states that this  sum is a topological invariant  equal (in this case)
to  unity \cite{Kurchan},  implying that  not  all TAP  states can  be
minima.  We shall resolve this apparent contradiction by showing that,
in  the thermodynamic  limit and  within the  BM solution,  the matrix
$\partial  G_i/\partial  m_j$ is  positive  {\em semi}-definite,  with
exactly one null eigenvalue.  For  $N$ large but finite, we argue that
each infinite-$N$ solution splits into  a pair consisting of a minimum
and an index-one saddle point,  both types of solution having the same
(extensive) complexity thereby rescuing the Morse theorem.

Recent work on spin-glass complexity \cite{CGPM,ACGP,CLPR1,ACGPT,CLPR2}  
exploits a  symmetry of the  action, the so-called  BRST supersymmetry
\cite{BRST}, that  enters the calculation of $N_s$.   If this symmetry
is unbroken,  the Morse theorem follows.  However,  thr solutions that
preserve  this symmetry have  been shown  \cite{CLPR2} to  violate the
convexity inequality  $\langle \ln N_s  \rangle_J \le \ln  \langle N_s
\rangle_J$,  where the brackets  represent disorder  averages, leading
Crisanti  et al.\  to  conjecture  that there  might  be no  extensive
complexity \cite{CLPR2}. The BM  solution breaks the BRST symmetry and
satisfies   all   physical   requirements,   provided   the   apparent
difficulties with the Morse theorem can be resolved. It is the purpose
of this  paper to present  such a resolution.  As a spin-off  from our
calculation,  we  note  that  the  marginal stability  of  TAP  states
provides a possible explanation for why they are  so difficult to find
numerically.

The free  energy (multiplied by $\beta  = 1/k_BT$) of  a TAP
state is given by \cite{TAP} 
\begin{eqnarray}
F &=& -\frac{\beta}{2}\sum_{i,j}J_{ij}m_im_j-\frac{N}{4}\beta^2(1-q)^2  
- N \ln 2 \nonumber  \\ 
&& + \sum_i\left(\frac{1}{2}\ln(1-m_i^2) + m_i\tanh^{-1}m_i\right),
\label{F}
\end{eqnarray}
where $q=(1/N)\sum_i m_i^2$, and the bonds  $J_{ij}$ are drawn from 
a Gaussian distribution of mean zero and variance $1/N$.

The TAP equations  are given by $G_i \equiv  \partial F/\partial m_i =
0$, for all $i=1,\ldots,N$, where
\begin{equation}
G_i = \tanh^{-1}m_i + \beta^2(1-q)m_i - \beta \sum_j J_{ij}m_j.
\label{G}
\end{equation}

Using $G_i=0$, the free energy $F$ can be rewritten as a sum of 
single-site  terms,  $F = \sum_i f_1(m_i)$, where
\begin{eqnarray}
f_1(m)  &  =  &  -\ln  2 -  (1/4)\beta^2(1-q^2)  +  (m/2)\tanh^{-1}m
\nonumber \\ && \hspace*{3cm} + (1/2)\ln(1-m^2).
\end{eqnarray}
The number  of solutions (per unit free-energy range) with free energy 
per spin $f=F/N$ is given by 
\begin{eqnarray}
N_s(f)   &   =  &   N^2   \int_0^1   dq  \int_{-1}^{1}   \prod_i(dm_i)
\delta\left(Nq   -  \sum_i   m_i^2\right)  \nonumber   \\   &&  \times
\delta\left(Nf  - \sum_i f_1(m_i)\right)  \prod_i\delta(G_i)|\det {\bf
A}|,
\label{NS}
\end{eqnarray}
where ${\bf A}$ is the Hessian matrix,
\begin{eqnarray}
A_{ij}&=&\partial G_i/\partial m_j=\partial^2 f/\partial m_i\partial m_j
\nonumber \\
&=&\!\left[\frac{1}{1-m_i^2}\!+\!\beta^2(1-q)\right]\!\delta_{ij}\! 
- \!\beta J_{ij}\!-\!\frac{2\beta^2}{N} m_im_j.
\label{A}
\end{eqnarray}

The final term in Eq.\ (\ref{A}) is $O(1/N)$ and was omitted in BM.  
It does not contribute to the extensive part of  the complexity, only 
to the prefactor of the exponential in  the  relation $\langle  N_s(f)
\rangle_J  \sim \exp[N\Sigma(f)]$.   It has,  however, the  form  of a
projector and may play an important role in the eigenvalue spectrum of
the  matrix ${\bf A}$,  as emphasized  by Plefka  \cite{Plefka82a}. In
particular, it may determine the  sign of $\det{\bf A}$. We shall show
that,  for $N  \to \infty$,  the projector  term splits  off  a single
isolated  eigenvalue  from  the  continuous  spectrum  of  ${\bf  A}$.
Furthermore, the  continuous part contains  only positive eigenvalues,
while  the  isolated  eigenvalue  is  a null  eigenvalue  outside  the
continuum.

Eq.\ (\ref{NS}) is  the common starting point for  all calculations of
$N_s(f)$. Here we focus on the configuration average (sometimes called
the  ``annealed  average''  or  ``white  average''),  $\langle  N_s(f)
\rangle_J$,  over realizations  of the  disorder. However,  it  is not
straightforward  to  do  this  while  retaining  the  modulus  on  the
determinant.  Dropping  the modulus, the calculation  can be completed
and the result takes the form \cite{BM80}
\begin{eqnarray}
\frac{1}{N}\ln  \langle  N_s(f)  \rangle_J   &  =  &  -\lambda  q  -uf
-(B+\Delta)(1-q)\nonumber \\ && +(B^2 - \Delta^2)/2\beta^2 + \ln I,
\label{N_s}
\end{eqnarray}
where $I$ is a function of the parameters $\lambda$, $q$, $u$, $B$ and
$\Delta$ and is defined by the integral
\begin{eqnarray}
I  & = &  \int_{-1}^{1}\frac{dm}{\sqrt{2\pi P}}\left(\frac{1}{1-m^2}+B
\right)  \exp \bigg[\lambda  m^2  + uf_1(m)  \nonumber \\  &&
\hspace{2cm}  -\frac{(\tanh^{-1}m - \Delta m)^2}{2P} \bigg],
\label{I}
\end{eqnarray}
where $P=\beta^2 q$.  The left-hand side of (\ref{N_s}) is the
complexity  or, more properly,  the ``annealed  complexity''.  The
parameters    $\lambda$,\ldots,$\Delta$    originally   entered    the
calculation  as integration  variables:  $\lambda$ and  $u$ appear  in
auxiliary integrations  that relax  the delta function  constraints on
$q$ and  $f$ respectively, while the other  parameters were introduced
via Hubbard-Stratonovich transformations that  reduce the problem to a
single-site  problem. Full  details can  be found  in  \cite{BM80} and
\cite{CGPM}. The resulting  five-dimensional integral can be evaluated
(for $N \to  \infty$) by the method of steepest  descents, so the five
parameters take  values corresponding to the  appropriate saddle point
in the  five-dimensional space.  Note that a  calculation of  the {\em
total} number of solutions, independent of their free energy, requires
setting $u=0$.

It is  straightforward to derive the five  saddle-point equations from
Eqs.\  (\ref{N_s}) and (\ref{I}).  The same  five equations  appear in
\cite{BM80}  and  \cite{CGPM}.  Equivalent equations  have  also  been
derived by De  Dominicis et al.\ \cite{DGGO}. The  equations admit the
solution  $B=0$,  and  this  is  the solution  adopted  in  all  three
papers. The  differences between the  various treatments arise  in the
solution of the remaining four equations.

Consider first the  case $u=0$, corresponding to a  calculation of the
total number of solutions. Cavagna et al.\ \cite{CGPM} (CGPM) note that
the BM solution apparently violates the Morse theorem, and propose a 
new BRST-symmetric solution that gives vanishing complexity for $u=0$. 
As  $u$  is decreased,  states  of lower  free energy are  selected and,  
within their solution, CGPM find that there exists a threshold $f$ below 
which $\ln \langle N_s(f) \rangle_J$   is   nonzero.   Unfortunately,  
however,   an   important inequality, 
$x_p  \equiv 1  - (\beta^2/N)\sum_i(1-m_i^2)^2 \ge  0$, is
violated in  the CGPM solution, rendering  it unphysical \cite{CLPR1}.
The condition $x_p \ge 0$ is necessary for the internal consistency of
the  TAP  equations  \cite{Plefka82a,Plefka82b}.  This  inequality  is
satisfied  by  the  BM  solution  \cite{CLPR1}.  The  BM  solution  is
internally  consistent, therefore, provided  one can  demonstrate that
the matrix ${\bf A}$ is positive definite, guaranteeing the positivity
of the determinant and justifying  the replacement of $|\det {\bf A}|$
by $\det {\bf A}$ in  the calculation, and provided one can understand
the  apparent  violation  of  the  Morse  theorem  that  ensues.   The
remainder of the Letter is devoted to these subtle points.

We first  rewrite Eq.\  (5) in the  form $A_{ij} =  (X^{-1})_{ij} +
(2\beta^2/N)m_im_j$, in  which the  projector term has  been separated
off.   The  matrix  ${\bf   A}^{-1}$  is  the  susceptibility  matrix,
$(A^{-1})_{ij} = \partial m_i/\partial  h_j$, giving the response to a
site-dependent   external  field,  and   $X_{ij}$  is   the  $O(1)$
contribution     to     it.      The    eigenvalue     spectrum     of
${\bf X}^{-1}$ can be  obtained using either Pastur's
theorem \cite{Pastur} or the `locator expansion' of ref.\ \cite{BM79}.
In the  limit $N  \to \infty$, the  spectrum consists of  a continuous
band of positive eigenvalues for  both $x_p>0$ and $x_p<0$ (though, as
noted, the  TAP equations  themselves become unphysical  for $x_p<0$),
and the left  edge of the band reaches zero only  for $x_p=0$. For the
BM      solution      $x_p>0$      so     all      eigenvalues      of
${\bf X}^{-1}$ are positive.  When the projector term
is  included,  an  isolated  eigenvalue,  outside the  main  band,  is
produced. Using the eigenvectors of ${\bf X}^{-1}$ as
a basis,  it it easy to  show \cite{details} that this eigenvalue  has 
a non-negative value provided the inequality
\begin{equation}
1 \ge \frac{2\beta^2}{N} \sum_{ij} m_i X_{ij} m_j = 2\beta^2 H
\label{H}
\end{equation}
is satisfied, where the final  equality defines $H$. The same result 
can be obtained using the variational trial function 
$v_i = \sum_j X_{ij}m_j$ for the eigenvector of ${\bf A}$ with 
smallest eigenvalue, i.e.\ 
$\lambda_{\rm min} \le \sum_{i,j}v_iA_{ij}v_j/\sum_i v_i^2 
\propto (1-2\beta^2 H)$. If the inequality (\ref{H}) becomes an equality, 
the isolated eigenvalue $\lambda_{min}$ has the value zero, and the 
variational eigenfunction becomes exact. A variant of the inequality 
(\ref{H}) was derived earlier \cite{Plefka82a}, with only the diagonal 
terms, $i=j$, appearing on the right. The off-diagonal terms were missing 
due to the use of Pastur's theorem outside its range of validity \cite{Owen}. 

The quantity $H$ in Eq.\ (\ref{H}) can be computed as follows. 
We introduce an additional factor of unity, expressed as  
\begin{equation}
1\! = \!\frac{1}{\sqrt{\det{\bf X}}} 
\int_{-\infty}^\infty \prod_i\left(\frac{d\phi_i} {\sqrt{2\pi}}\right)
\exp\left(-\frac{1}{2}\sum_{i,j}\phi_i (X^{-1})_{ij} \phi_j\right)\!,
\label{phi}
\end{equation}
in the integrand of Eq.\ (\ref{NS}), and obtain $H$ from  
$H = (1/N) \sum_{ij} \langle m_im_j \langle \phi_i \phi_j \rangle_\phi
\rangle_m$, where the averages $\langle \ldots \rangle_\phi$ and 
$\langle \ldots \rangle_m$ are over the variables $\{\phi_i\}$ and $\{m_i\}$ 
respectively. The weight function for the $\phi_i$ integrals is given the 
integrand in Eq.\ (\ref{phi}), while for the $m_i$ integrals it is given by 
the integrand in Eq.\ (\ref{NS}).    

After a straightforward but lengthy calculation one finds \cite{details}
\begin{equation}
H = \frac{A_3 q^2}{(q-A_1)^2 + A_3[\beta^2q(1-q)-A_2]}
\end{equation}
where
\begin{eqnarray}
A_1 & = & \langle (1-m^2)m(\ath m -\Delta m) \rangle \\
A_2 & = & \langle (1-m^2)(\ath m -\Delta m)^2 \rangle \\
A_3 & = & \langle m^2(1-m^2) \rangle
\end{eqnarray}
and  the averages  are now  over the  weight function  given  by the
integrand  in Eq.\  (\ref{I}).  Carrying out  the required  integrals
numerically (with $B=0$ as usual) one obtains a remarkable result: the
quantity $2\beta^2 H$ is unity for  all temperatures $T < T_c$ and all
values of the free energy per spin, $f$, i.e. the inequality (\ref{H})
is  satisfied  as  an equality.  We  feel  it  should be  possible  to
demonstrate  this  result  analytically  but  thus  far have  not
succeeded.

This  result, that  in the  thermodynamic  limit there  is always  one
exactly zero  eigenvalue, but  no negative eigenvalue,  is the  key to
resolving all  the puzzles surrounding this  problem. First, $\det{\bf
A}$ vanishes,  so the prefactor  of the exponentially large  number of
TAP states is, for $N \to\infty$, exactly zero, in accordance with the
result of  Kurchan \cite{Kurchan} and its extension  to general values
of  $u$ (the variable  conjugate to  $f$) \cite{CLPR2}.   However, the
exponential  itself diverges  for $N  \to \infty$,  so the  product of
exponential and prefactor is not  defined in this limit. To make sense
of it, one has to keep  $N$ large but finite. The result, confirmed by
numerical studies, is that the  zero eigenvalue is shifted, for finite
$N$, to  a small  positive or negative  value, corresponding to  a TAP
minimum or to  a saddle of index one respectively.  The shift would be
expected to  be of order  $1/\sqrt{N}$ \cite{details}. No  examples of
more  than one  negative eigenvalue  were found.   Furthermore,  for a
given sample the  two types of solution typically  occur together as a
closely related pair, in a sense we will clarify below. The extrema of
the finite-$N$ TAP  free energy are therefore dominated  by minima and
index-one saddles.

This picture can  be further clarified by constructing a fictive 
free-energy function
\begin{equation}
F_q   =  \tilde{F} + \frac{\beta^2}{2}(1-q)\left(\sum_im_i^2  - Nq\right),
\label{Fq}
\end{equation} 
where  $\tilde{F}$ is a function of the $m_i$ and $q$. It is given by  
Eq.\  (\ref{F}), but with $q$ regarded as an independent variable, 
unrelated to the $m_i$, i.e.\ $F_q$ is a function  
of the $N+1$ variables $m_1,\ldots,m_N,q$, whereas the original TAP free 
energy $F$ depends only  on the $N$  variables $m_1,\ldots,m_N$ (with $q$  
{\em defined}  as $q=(1/N)\sum_i  m_i^2$). One readily verifies that the 
stationarity equations for $F_q$ reproduce the TAP equations: 
$\partial  F_q/\partial  m_i  =  G_i  = 0$.  However,  for  these  new 
equations, the quantity $Q \equiv (1/N)\sum_i m_i^2$ is in general not equal  
to the  parameter $q$  appearing  in the  equations. The additional   
stationarity  equation,   $0   =  \partial F_q/\partial q = 
(\beta^2/2)(Nq - \sum_i m_i^2)$ forces $Q=q$ at stationary points 
in the full $(N+1)$-dimensional space. The free-energy functions $F$ and 
$F_q$ have, therefore, the same stationary points and the same values at 
these points. By formally eliminating the variables $m_i$, one can obtain 
the function $F_q(q)$ as a function of the single variable $q$. Its first 
derivative is $dF_q/dq = (\beta^2/2)(Nq - \sum_i m_i^2)$, where the $m_i$ 
are implicit functions of $q$ through the TAP equations.  In practice, 
of course, there will be exponentially many TAP solutions, 
$\langle N_s \rangle_J \sim \exp[N\Sigma_q]$, for each fixed value of $q$. 
Their number can be calculated from the same equations, (\ref{N_s}) and 
(\ref{I}) as before, but with $\lambda=0$, since $q$ is no longer 
constrained to equal $(1/N)\sum_i m_i^2$, and $u=0$ since $f$ is not fixed. 
The functions $F_q(q)$ and $Q(q)$, however, are {\em self-averaging} and 
therefore well-defined, being determined by averages of the appropriate 
functions of $m$, e.g.\ $Q=\langle m^2 \rangle$, where the weight function 
for the averages is the integrand of Eq.\ (\ref{I}), with $\lambda=0=u$. 
  
\begin{figure}
\includegraphics[width=\linewidth]{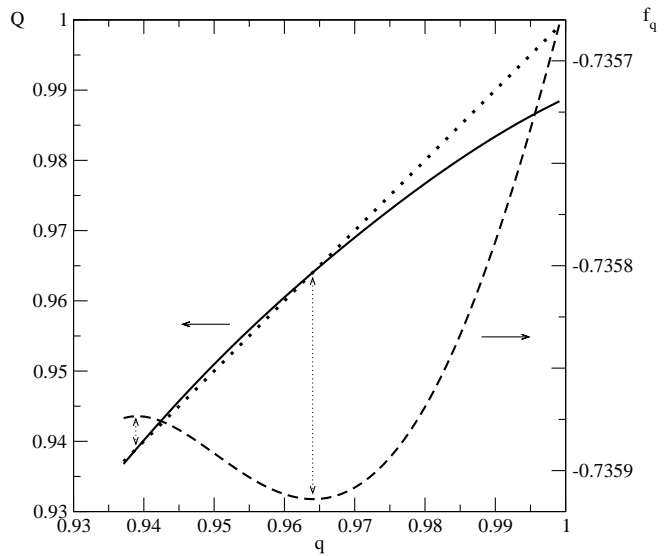}
\caption{\label{figure1} The functions $Q(q)$ (continuous line) and 
$f_q(q)$ (dashed line) defined in the text. Physical states occur where 
$Q(q)$ crosses the dotted line $Q=q$. The date were obtained using 
$N=200$ spins at a temperature $T=0.2$.} 
\end{figure}

We have  solved these  TAP-like equations numerically  for a  range of
$q$, and  determined the corresponding values of  $Q$ and $f_q=F_q/N$.
An example is  show in Figure 1. First a solution  of the standard TAP
equations (i.e.\ with $q=(1/N)\sum_i  m_i^2$) was found, and solutions
for  other $q$-values  were  generated iteratively  from the  previous
value,  starting  from  the  TAP solution.   The  iterative  procedure
typically fails to  coverge when $q$ becomes too  small.  The physical
solutions in  Figure 1  are the two  points where the  function $Q(q)$
intersects the  line $Q=q$. They  correspond to turning points  of the
function  $f_q(q)$.    The  solution   with  the  larger   $q$  always
corresponds to a minimum of $F$,  the other solution to a saddle point
of  index  one.  The  difference  vector,  $\delta  m_i$, between  the
solutions typically  has a large overlap with  the eigenvector, $e_i$,
of  ${\bf A}$ with the smallest  eigenvalue: $\sum_i  e_i \delta
m_i/[\sum_i e_i^2 \sum_j (\delta  m_j)^2]^{1/2} \approx 0.1 - 1$, with
a typical value around 0.5. This shows that one moves from the minimum
to the saddle point by moving roughly in the direction of the isolated
``soft  mode''. This agrees with our expectation based on the relation 
$dm_i/dq = \beta^2 \sum_j X_{ij}m_j$, which follows from the
TAP equation. Recall that $v_i  = \sum_j X_{ij}m_j$ becomes, for $N
\to \infty$, the null eigenfunction  of ${\bf A}$, so $dm_i/dq \propto
v_i$ in  this limit. The minimum  and the saddle  point will 
coalesce as  the small eigenvalue  tends to zero with  increasing $N$,
and  the  two  turning  points  of  $f_q(q)$ will  merge  to  form  an
inflection  point. One can  see this  formally by  differentiating the
relation   $df_q/dq   =  (\beta^2/2)[q-(1/N)\sum_i m_i^2]$ to obtain
\begin{equation}
\frac{d^2f_q}{dq^2}=\frac{\beta^2}{2}\left(1-\frac{2\beta^2}{N}
\sum_{ij}m_iX_{ij}m_j\right) = 0\ .
\end{equation} 

It  is important  to  recall  that the  isolated  eigenvalue of  order
$1/\sqrt{N}$ does not  enter the result for the  extensive part of the
complexity, because the projector  term in Eq.\ ({\ref A}) responsible
for it is  $O(1/N)$ and drops out of the  complexity at leading order.
The upshot is that the BM  calculation, in which the projector term is
neglected,  counts minima  and index-one  saddles, both  with positive
sign, since without  the projector term the Hessian  matrix is, for $N
\to  \infty$, positive  definite.  We  have shown  that  including the
projector  produces one  null eigenvalue  in the  thermodynamic limit,
i.e.\  the prefactor  in the  calculation of  $\langle  N_s \rangle_J$
vanishes  as  required  by  exact analysis  \cite{Kurchan,CLPR2}.  For
finite  $N$, however,  the marginally  stable states  become  pairs of
minima and index-one saddles.

This  suggests a scenario  in which  the complexities  associated with
minima and  index-one saddles  are extensive and  equal, and  no other
solutions are possible in the  limit $N \to \infty$ except the trivial
solution  $m_i=0$.   The  Morse  theorem  would  then  be  identically
satisfied.   Saddles of  index  greater  than one  can  only occur  if
${\bf X}^{-1}$  develops   negative  eigenvalues  for
finite $N$. Then  one would need to calculate  the probability, $p_k$,
to have  $k$ negative eigenvalues. This probability  would be expected
to be exponentially small in  $N$, $p_k \sim \exp(-a_k N)$.  Just such
a computation has been carried out in the (simpler) $p$-spin spherical
spin-glass  \cite{CGP}. If  the  coefficients $a_k$  are large  enough
(larger than the computed complexity of minima and index-one saddles),
the  complexity  of  index-$k$  saddles  will be  negative  and  these
higher-order saddles will not contribute to the Morse sum.

It has been argued \cite{CLPR2} that numerical studies \cite{Plefka02}
of  the TAP equation  are difficult  to reconcile  with the  BM theory
since  the  range of  observed  free  energies  is much  smaller  than
expected,   and  the  measured   value  of   $x_p$  is   smaller  than
predicted.  However,  the  numerical  solutions are  obtained  from  a
dynamical  algorithm and it  is known  that such  algorithms typically
generate solutions of a given (algorithm-dependent) free energy rather
than  generating  them  uniformly  from  the  underlying  distribution
\cite{NewmanStein}.  Moreover,  the free  energy of such  solutions is
typically  significantly lower than  the dominant  free energy  of the
underlying  distribution,  implying  that  the quantity  $x_p$  (which
decreases  as  $f$  decreases)   will  indeed  be  smaller  for  these
dynamically generated states than  for the dominant states selected by
the BM solution. 

We  conclude that the  BM theory remains a viable 
candidate theory of spin-glass complexity. 

TA acknowledges financial support from the European Community's Human 
Potential Programme under contract HPRN-CT-2002-00307, DYGLAGEMEM.

\end{document}